\documentclass[prl,twocolumn,showpacs,aps]{revtex4}
\usepackage{graphics}
\usepackage[sort&compress]{natbib}
\newcommand{\rb}{{\mathbf{r}}}
\newcommand{\Vb}{{\mathbf{V}}}

\newcommand{\mPhi}{{\mit\Phi}}

\begin{document}
\title{Topological quantum correction to an atomic ideal gas law as
a dark energy effect}
\author{Eugene V. Kholopov}
\affiliation{A.V. Nikolaev Institute of Inorganic Chemistry, Siberian 
Branch, Russian Academy of Sciences, 630090 Novosibirsk, Russia}
\email{kholopov@niic.nsc.ru}
\affiliation{Novosibirsk State University, 630090 Novosibirsk, Russia}

\maketitle

{\bf The traditional ambiguity about the bulk electrostatic potentials in 
crystals is due to the conditional convergence of Coulomb series \cite
{Tosi64,Harr75}. The classical Ewald approach \cite{Ewal21,Born54} turns 
out to be the first one resolving this task as consistent with a 
translational symmetry \cite{Khol04,Khol07,Khol10}. The latter result
appears to be directly associated with the thermodynamic limit in 
crystals \cite{Kho06a}. In this case the solution can also be obtained 
upon direct lattice summation, but after subtracting the mean Bethe 
potential \cite{Khol04,Kho06b}. As shown \cite{Kho06b}, this effect is 
associated with special periodic boundary conditions at infinity so as to
neutralize an arbitrary choice of the unit-cell charge distribution.
However, the fact that any additional potential exerted by some charge
distribution must in turn affect that charge distribution in equilibrium
is not discussed in the case at hand so far. Here we show that in the
simplest event of gaseous atomic hydrogen as an example, the
self-consistent mean-field-potential correction results in an additional
pressure contribution to an ideal gas law. As a result, the corresponding
correction to the sound velocity arises. Moreover, if gas in question is
not bounded by any fixed volume, then some acceleration within that
medium is expected. Addressed to the Friedman hypersphere \cite{Frie22}, 
our result may be interesting in connection with the accelerating
Universe revealed experimentally \cite{Ries98,Schm98,Perl99} and discussed 
intensively \cite{Peeb03,Cher08,Luka08,Bart10}.}

\begin{figure}[t]
\resizebox{0.95\hsize}{!}{\includegraphics{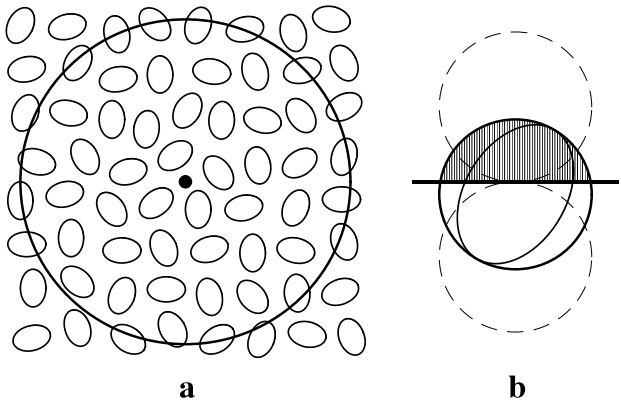}}
\caption{{\bf An electrostatic potential in a uniform isotropic medium is determined by two steps.} The global one shown in panel (a) proposes the account of atomic contributions (small ovals) inside an ideal sphere (the large circle) centred at a reference point (the small filled circle) and of radius $R_0$ tending to infinity. Then the boundary effect associated with truncated ovals arises. In the case of an individual oval shown in panel (b) this effect includes the averaging over orientations transforming that heavy oval to a sphere (the solid circle) of radius $R$. This sphere is further truncated by the horizontal line exhibiting the boundary surface element. The charge containing in the outer part of the atomic sphere (the hatching area) is suggested to be mapped onto the surface element for neutrality. All possible truncations are restricted schematically by the dashed circles.} \label{Fig1}
\end{figure}

Let us consider a uniform rarefied gaseous medium built up of 
electrically neutral atomic or molecular objects of one species, for 
simplicity, described by a charge distribution $\rho_{\rm ini}(\rb)$ and 
called as 'atoms' for definiteness. If the shape of $\rho_{\rm ini}(\rb)$ 
is not spherically symmetric, then all possible orientations of 
$\rho_{\rm ini}(\rb)$ are expected in the ensemble under consideration, 
as shown in Fig. \ref{Fig1}a. This ensemble is assumed to be uniform and 
isotropic. Then the electrostatic potential at any reference point may be 
thought of as an effect exerted by a spherical volume centred at the 
reference point, as shown in Fig. \ref{Fig1}a too. It is important that 
the multipole contribution to that potential, if happens, is expected to 
be rather short-range and so vanish on the remote spherical boundary. 
Another contribution as a boundary effect takes place if we propose that 
the boundary truncates atoms in the boundary region. If the radius $R_0$ 
of that spherical boundary is large enough, then all possible truncations 
of atoms are available. Furthermore, the boundary element truncating an 
atom can be treated as a plane one in this case. In order to conserve the 
electrical neutrality, the outer part of every atomic charge distribution 
is suggested to be mapped onto the truncating plane. As a result, the 
averaging over atomic orientations leads to a spherically symmetric 
effective charge distribution $\rho(r)$, where $r = |\rb|$. The further 
averaging over atomic positions and mapping depicted in Fig. \ref{Fig1}b 
gives rise to the effect of double layer \cite{Kho06b}. It is significant 
that the potential effect at the central reference point exerted by the 
double layer arising on the boundary surface turns out to be independent 
of a concrete set of truncated atoms and so the final effect remains the 
same if the large radius $R_0$ is changed anyhow. It implies that the 
potential in question is of topological nature. It is easy to show
\cite{Kho06b} that this potential arising in the interior of the ensemble 
as a correction takes the form
\begin{equation}\label{Aq1}
\mPhi_{\rm top}=\frac{2\pi}{3v}\int\limits_Vr^2\rho(r)d\rb ,
\end{equation} 
where $v$ is the volume per atom in question. As pointed our earlier \cite
{Kho06b}, this result can also be thought of as a compensation of the 
Bethe mean potential.

In order to understand the self-consistent character of 
$\mPhi_{\rm top}$, we consider the simplest case of $\rho(r)$ associated 
with the hydrogen atom described by the electron wave function 
$\psi(r)\propto\exp(-\alpha r)$, where $r$ is a distance from the central 
proton. Due to 
the spherical symmetry of $\psi(r)$, one can see that the normalized 
electron charge distribution necessary for (\ref{Aq1}) is equal to 
$\rho(r)=-(e\alpha^3/\pi)\exp(-2\alpha r)$, where $e$ is the elementary 
charge. With taking potential (\ref{Aq1}) into account, the electron 
contribution to the ground-state energy of the hydrogen atom with a 
certain electron spin projection can be written down as
\begin{equation}\label{Aq2}
E_{\rm mod}=\frac{\hbar^2\alpha^2}{2m_{\rm e}}-\alpha e^2
+\frac{\pi e^2}{\alpha^2v} .
\end{equation} 
Here $\hbar$ is Planck's constant, $m_{\rm e}$ is the electron mass, the 
last term on the right-hand side of (2) corresponds to the foregoing 
effect of $\mPhi_{\rm top}$, but operating on the electron part of the 
hydrogen atom only. This is a subtle point of the present consideration 
where the self-consistency of the configuration of the electron subsystem 
alone is studied. Moreover, a half of the effect of $\mPhi_{\rm top}$ 
associated with a particular spin projection seems to be plausible to be 
taken into account. An equilibrium value of $\alpha$ is then obtained 
from the condition that $E_{\rm mod}$ be a minimum. As a result, we derive
\begin{equation}\label{Aq3}
\frac{\hbar^2\alpha}{m_{\rm e}}-e^2-\frac{2\pi e^2}{\alpha^3v}=0 
\end{equation} 
so that $\alpha$ becomes a function of $v$ now.

Here we are interested in the case of rarefied gas where the value of $v$ 
is large enough so that the value of $\alpha$ is very close to its 
conventional magnitude of $\alpha_0=m_{\rm e}e^2/\hbar^2$ in agreement 
with formula (\ref{Aq3}). The case of condensed matter will be discussed 
elsewhere. It is important that connection (\ref{Aq3}) takes place even 
if we come back to the general atomic energy $E$ specified by relation 
(\ref{Aq2}), where the last term should be omitted as applied to a 
neutral atom as a whole. In particular, an additional pressure effect can 
be expected therefrom in the form modifying an ideal gas law:
\begin{equation}\label{Aq4}
P=P_{\rm top}+\frac{k_{\rm B}T}{v} ,
\end{equation} 
where $k_{\rm B}$ is the Boltzmann constant, $T$ is the temperature,
\begin{eqnarray}\label{Aq5}
P_{\rm top}&=&-\frac{\partial E}{\partial v}=
-\frac{\partial E}{\partial\alpha}\frac{d\alpha}{dv}=
\frac{4\pi^2m_{\rm e}e^4}{\alpha^6v^3\hbar^2}
\Bigl(1+\frac{6\pi m_{\rm e}e^2}{\alpha^4v\hbar^2}\Bigr)^{-1}\nonumber\\
&\approx&\frac{4\pi^2m_{\rm e}e^4}{\alpha^6v^3\hbar^2} .
\end{eqnarray} 
The last relation corresponds to the limit of large $v$. In this case 
$P_{\rm top}v_0^3\approx1.1615\cdot10^{15}$ Pa, where the dimensionless 
$v_0$ is measured in atomic units.

It is interesting to note that both the terms on the right-hand side of 
equation (\ref{Aq4}) are of kinetic nature. Indeed, while the last term 
there is associated with the original statistical Maxwell distribution, 
the first one is connected with the disturbance of the kinetic intratomic 
energy in form (\ref{Aq2}). 

The additional pressure effect obtained above immediately leads to the 
existence of the corresponding infrasound that is especially pronounced 
in the limit of zero temperature. In order to derive it, we introduce the 
corresponding mass density $\mu=m_{\rm p}/v$, where $m_{\rm p}$ is the 
proton mass. Now we cast the foregoing value of $P_{\rm top}$ in terms of 
$\mu$. The sound velocity of interest $c_{\rm s}$ in the limit of large 
$v$ is then represented as
\begin{equation}\label{Aq6}
c_{\rm s}^2=\frac{\partial P_{\rm top}}{\partial\mu}
\approx\frac{12\pi^2}{\alpha^6v^2}\frac{m_{\rm e}e^4}{m_{\rm p}\hbar^2}
=\frac{3.09\cdot10^{11}}{v_0^2}\Bigr(\frac{\rm m}{\rm s} \Bigl)^2 .
\end{equation} 

It is important that the effect mentioned above takes place only when gas 
in question is fixed anyhow in space. Otherwise, the value of $v$ tends 
to increase. In order to describe this problem we introduce Cartesian 
parameters of unit length $r_x$, $r_y$ and $r_z$, so that $r_xr_yr_z=v$. 
On the other hand, the derivatives of those length parameters with 
respect to time $t$ represent the corresponding velocity components 
$V_\beta=dr_\beta/dt$, providing that $V_x^2+V_y^2+V_z^2=\Vb^2$. As a 
result, in the non-relativistic limit the essential part of the local 
energy can be written as $E_{\rm loc}=m_{\rm p}\Vb^2/2+E$, providing that 
every volume $v$ is specified by one atom with the proton mass 
$m_{\rm p}$. The conservation of this value implies that 
$dE_{\rm loc}/dt=0$. With taking equations (\ref{Aq3}) and (\ref{Aq5}) 
into account, the direct differentiation results in the following 
expression
\begin{equation}\label{Aq7}
\sum\limits_\beta V_\beta\Bigl[m_{\rm p}\frac{dV_\beta}{dt}
-\frac{4\pi^2m_{\rm e}e^4}{\alpha^6v^2r_\beta\hbar^2}
\Bigl(1+\frac{6\pi m_{\rm e}e^2}{\alpha^4v\hbar^2}\Bigr)^{-1}\Bigr]=0 .
\end{equation} 
At this stage of consideration it is convenient to consider the values 
$V_x$, $V_y$ and $V_z$ as mutually independent. It implies that each of 
the three expressions in the square brackets in the summand of 
(\ref{Aq7}) be zero. On the other hand, in the isotropic case of interest 
we believe that all the unit-cell parameters increases in the same manner 
so that they can be described in the form $r_x=r_y=r_z=x_0Q$, where $Q$ 
is a variable dimensionless parameter. Substituting the latter relation 
into formula (\ref{Aq7}), where each square brackets gives the same 
result, we get
\begin{equation}\label{Aq8}
\ddot Q=\frac{a}{Q^4(Q^3+b)} ,
\end{equation} 
where every dot over Q stands for the derivative with respect to time, 
the parameters
\begin{equation}\label{Aq9}
a=\frac{4\pi^2m_{\rm e}e^4}{\alpha^6x_0^8m_{\rm p}\hbar^2}\qquad
\mbox{and}\qquad b=\frac{6\pi m_{\rm e}e^2}{\alpha^4x_0^3\hbar^2}
\end{equation} 
are suggested to be independent of $Q$ that is right at least as 
$\alpha\to\alpha_0$. In this case equation (\ref{Aq8}) can be easily 
integrated and we obtain
\begin{equation}\label{Aq10}
\bigl({\dot Q}\bigr)^2=\frac{2a}{3b^2}\Bigl[b\Bigl(1-\frac{1}{Q^3}\Bigr)
+\ln\frac{Q^3+b}{Q^3(1+b)}\Bigr] ,
\end{equation} 
where an arbitrary constant of integration is chosen here so that ${\dot 
Q}=0$ at $Q=1$, for convenience. It implies that $Q=1$ is the initial 
value corresponding to a certain value of $x_0$ describing the state of 
gas before its expansion.

\begin{figure}[t]
\resizebox{0.6\hsize}{!}{\includegraphics{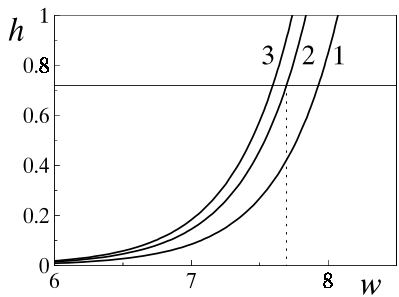}}
\caption{{\bf The dimensionless Hubble parameter $h$ in dependence on the 
initial gas density $n_{\rm ini}$ in units of $10^w$ atoms/m$^3$.} Curves 
1, 2 and 3 correspond to the final gas density equal to 1, 5 and 10 
atoms/m$^3$, respectively. The thin horizontal line describes the present 
value of the Hubble constant adopted as $h=0.719$. For curve 2 it 
corresponds to $n_{\rm ini}\approx5\cdot10^7$ atoms/m$^3$, as shown by 
the vertical dotted line.}\label{Fig2}
\end{figure}

It is surprising that relation (\ref{Aq10}) can be addressed to the 
cosmological problem. Indeed, the expansion of $v$ as a local effect 
leading to relation (\ref{Aq7}) can happen in the most natural way on the 
surface of the Friedman three-dimensional hypersphere \cite{Frie22}, 
providing that the radius of that hypersphere increases. In this event 
the value ${\dot Q}/Q$ may be associated with the Hubble function \cite
{Bart10,Tamm08} $H_0$ dependent on the initial gas density. In terms of 
the dimensionless parameter $h=H_0/(100\;{\rm km}\;{\rm s}^{-1} {\rm 
Mpc}^{-1})$ the results are shown in Fig. \ref{Fig2}. According to 
(\ref{Aq10}), it implies that the expansion predicted here is not  
associated with the Big Bang. The initial gas density may thus be 
regarded as the threshold one above which any fluctuation of distinct 
nature can be expected.

The time of expansion up to a given $Q$ can in turn be obtained upon 
further appropriate integration of equation (\ref{Aq10}):
\begin{equation}\label{Aq11}
t=b\sqrt{\frac{3}{2a}}\int\limits_1^Q \Bigl[b\Bigl(1-\frac{1}{u^3}\Bigr)
+\ln\frac{u^3+b}{u^3(1+b)}\Bigr]^{-1/2}du .
\end{equation} 
Relation (\ref{Aq11}) is simplified in the particular limit of small gas 
concentrations. In this case we obtain the conventional relation 
\cite{Bart10} $t\approx H_0^{-1}$. In the particular event singled out in 
Fig. \ref{Fig2} the time of interest becomes equal to 13.6 Gyr in 
agreement with the evaluations known in the literature 
\cite{Peeb03,Bart10}.

Here we consider the self-consistent effect of the topological field of 
the Coulomb nature in its simplest form with only one configuration 
parameter $\alpha$. In general, the same effect still exists, but the 
consideration is expected to be more complicated.
\vspace{3ex}

\noindent
P.S. This paper was submitted to Nature (2012-05-06944), Nature Physics 
(NPHYS-2012-06-01221-T), Nature Chemistry (NCHEM-12060801-T) and Nature 
Materials (NM12061695-T).

\end{document}